# Surface Chemical Reactivity of Ultrathin Pd(111) Films on Ru(0001): Importance of Orbital Symmetry in the Application of the d-Band Model.

Xiangshi Yin,[1,2] Valentino R. Cooper,[2] Hanno H. Weitering,[1,2*] and Paul C. Snijders[2,1*]

*hanno@utk.edu
*snijderspc@ornl.gov

[1] Department of Physics and Astronomy, The University of Tennessee, Knoxville, TN 37996, USA
[2] Materials Science and Technology Division, Oak Ridge National Laboratory, Oak Ridge, TN 37831, USA

**ABSTRACT**
The chemical bonding of adsorbate molecules on transition-metal surfaces is strongly influenced by the hybridization between the molecular orbitals and the metal d-band. The strength of this interaction is often correlated with the location of the metal d-band center relative to the Fermi level. Here, we exploit finite size effects in the electronic structure of ultrathin Pd(111) films grown on Ru(0001) to tune their reactivity by changing the film thickness one atom layer at a time, while keeping all other variables unchanged. Interestingly, while bulk Pd(111) is reactive towards oxygen, Pd(111) films below five monolayers are surprisingly inert. This observation is fully in line with the d-band model prediction when applied to the orbitals involved in the bonding. The shift of the d-band center with film thickness is primarily attributed to shifts in the partial density of states associated with the $4d_{xz}$ and $4d_{yz}$ orbitals. This study gives an in-depth look into the orbital specific contributions to the surface chemical reactivity, providing new insights that could be useful in surface catalysis.

**INTRODUCTION**

Heterogeneously catalyzed reactions typically start with the adsorption of reactant molecules onto the catalyst surface, followed by dissociation, reaction, and desorption.[1] As described by the Sabatier principle[2] the catalytic efficiency is determined by the strength of the interaction between reactants and catalyst: if the interaction is too weak, reactants do not adsorb or dissociate, whereas if the interaction is too strong the dissociation or reaction products do not desorb; thereby poisoning the catalyst surface.[3-5] The interaction strength is directly related to the catalyst surface electronic structure. The underlying physics for transition metal catalysts has been described quite successfully by the *d*-band model.[6-8] In this model, the interaction strength is determined by the location of the catalyst's *d*-band center, relative to the Fermi level. Here, the interaction of the adsorbate levels with the metal valence band can be pictured in two steps. The molecular levels first hybridize with the broad *sp* band of the metal, producing a rather broad resonance slightly below the free adsorbate levels. In a second step this renormalized state will interact with the relatively localized metal *d*-band, forming well-separated bonding and antibonding states. The strength of the resulting molecule-surface bond is determined the position and filling of the unperturbed metal *d*-levels, which, in turn, affects the filling of these anti-bonding states.

Indeed, while controlled tuning of catalytic activity remains difficult, electronic structure manipulation of a catalyst's surface appears to be a viable strategy toward tuning and enhancing its catalytic activity. Typical approaches to this end include the introduction of active sites with a locally modified *d*-band structure via doping or alloying,[9,10] or by varying the dimensions of the catalyst, exposing lattice sites with different local symmetries at step edges and high-index crystallographic planes on catalyst nanoparticles.[10-13] However, these approaches not only vary the *d*-band filling, but also change the local lattice symmetry and thus the type of orbitals involved in chemical reactions taking place at these active sites. Hence, the altered chemical activities cannot be attributed solely to shifts of the *d*-band center, leaving it an experimental challenge to properly validate the basic premise of the *d*-band model.

In this paper, we exploit finite size effects in atomically thin epitaxial films of Pd(111) on Ru(0001) as a means to tune the Pd *d*-band structure without introducing foreign species or changing local symmetries. To correlate the size-dependent electronic structure changes with possible changes in chemical reactivity, we probe the saturation coverage of oxygen as a function of the Pd film thickness. Oxygen dissociatively adsorbs on both bulk Ru(0001) and Pd(111) surfaces at room temperature (RT)[14,15] with saturation coverages of 0.5 and 0.25 ML for Ru and Pd, respectively,[16-20] for low oxygen partial pressures ($< 5 \times 10^{-5}$ mbar). Our results reveal a surprisingly strong non-monotonic Pd film thickness dependence of the saturation coverage. In particular, the oxygen saturation coverage on two-to-five ML thick Pd(111) films is extremely small, indicating that the Pd-O bonding is very weak for those films.

While changes in chemical reactivity and sticking probability on epitaxial Pt films on Ru(0001) have been observed,[21] these observations were mostly attributed to a charge transfer from the film to the substrate.[22,23] Moreover, the sticking probability was observed to increase monotonically with increasing thickness from 1 to 10 monolayers (ML), as opposed to the non-monotonic behavior observed here. Our Density Functional Theory (DFT) calculations point to finite size effects in the Pd band structure as a function of film thickness as the main cause of these observations. The observed increase in the oxygen sticking probability of the Pd films beyond 5 ML appears to be consistent with the central premise of the *d*-band model as the energy of the *d*-states that are involved in oxygen binding shifts toward the Fermi energy, ultimately stabilizing at the energy of bulk Pd. However, a comparison of the calculated oxygen binding energies and *d*-band center for Pd film thicknesses below 1 ML as well as those of the bulk Ru(0001) surface, indicates that a single-parameter description of the *d*-band model fails here, and that orbital symmetries must be taken into consideration. Hence, the thickness dependent reactivity of Pd(111) thin films on Ru(0001) not only provides a 'clean' validation of the *d*-band model, but also indicates that the orbital nature of the d-states involved in the chemical bonding are very important in establishing trends in catalytic activity.

**EXPERIMENTAL AND THEORETICAL PROCEDURES**

*Experimental procedures* - Experiments were conducted in an ultra-high vacuum (UHV) system with a base pressure of $10^{-11}$ mbar. The system is equipped with a variable temperature scanning tunneling microscope (STM), a single-pass cylindrical mirror analyzer for Auger electron spectroscopy (AES, resolution 0.6% of the electron energy, ac bias modulation 1 V), low energy electron diffraction (LEED), and an ion sputter gun for in situ sample preparation and cleaning. Sample heating was achieved with electron-beam heating from the backside of the sample. The Ru(0001) crystal used in this study was cleaned by cycles of 500 eV Ne ion bombardment at $5 \times 10^{-5}$ mbar at room temperature and post-annealing up to 1100 °C. Pd thin films were deposited onto the Ru substrate by direct heating of Pd in a home-made tungsten wire-basket. The Pd film thickness was calibrated with STM and AES [Appendix A], and monitored using a quartz crystal monitor. Oxygen was introduced into the UHV chamber with a variable leak valve at partial pressures between $2\times10^{-8}$ to $4\times10^{-8}$ mbar, with the sample at room temperature during exposure. Different oxygen exposures were obtained by varying the duration of the exposure at constant pressure. The peak-valley amplitude of the derivative $KL_{23}L_{23}$ oxygen Auger peak was used as a measure of the surface oxygen concentration.

*Theoretical procedures* - All DFT calculations employed projector augmented wave potentials with the generalized gradient approximation for exchange and correlation, as implemented in the Vienna Ab Initio Simulation Package (VASP v .5.3.3). A plane wave cutoff of 400 eV and an 8×8×1 Monkhorst k-point mesh were used for all slab calculations. The total energies were converged to $10^{-6}$ eV. We studied $n$ atomic Pd(111) layers on six Ru(0001) substrate layers, where $n$ = 0.75, 1, 2, 3. Calculations for these hybrid slabs assumed the theoretical in-plane lattice constant of 2.706 Å of bulk Ru(0001) which is in excellent agreement with the experimental value of 2.71 Å. Our theoretical lattice constant of 3.952 Å for bulk Pd is also in good agreement with the experimental value of 3.89 Å. Slab calculations for pure Ru(0001) and Pd(111) contained six atomic layers and were performed at their respective theoretical in-plane lattice constants. For all calculations, atoms were relaxed while keeping the in-plane lattice constant fixed, until the Hellman-Feynman forces on each atom were less than 0.01 eV/Å. For all slabs we consider a 2×2 in-plane periodicity with at least 16 Å of vacuum in between the slabs to ensure that there are no interactions between the top and bottom of the slabs. This surface periodicity results in four metal atoms per surface unit cell. This choice is motivated by the experimental saturation coverage of oxygen on Pd(111) of ¼ ML where the oxygen atoms form an ordered 2×2 structure. For completeness we also explored ½ ML coverage. For all coverages, oxygen atoms were assumed to adsorb in the fcc threefold hollow sites.[24,25]

The Gibbs free energy of adsorption, $\Delta G$ is computed as follows:

$$\Delta G(T,p) = E_{ads-surf} - E_{bare-surf} - N_O \mu_O(T,p) + ST + E_{vib} \quad (1)$$

here $E_{ads-surf}$ and $E_{bare-surf}$ are the DFT total energies at $T$ = 0 for the oxygen covered and bare surfaces, respectively. $N_O$ is the number of adsorbed O atoms per 2×2 supercell; $N_O$ = 1 and 2 at ¼ ML and ½ ML oxygen coverage, respectively. $S$ and $E_{vib}$ are the configurational and vibrational entropy contributions, respectively, which were both determined to be relatively small (< 50 meV) and thus ignored for the purpose of this paper. $\mu_O$ is the oxygen chemical potential at temperature $T$ and pressure $p$, given by:[26]

$$\mu_O(T, p_{O_2}) = \frac{1}{2}\left[E_{O_2} + \tilde{\mu}_{O_2}(T, p^0) + k_B T \ln\left(\frac{p_{O_2}}{p^0}\right)\right] \quad (2)$$

Here, $E_{O_2}$ = 4.88 eV is the total energy for an oxygen molecule at 0 K obtained from spin-polarized DFT calculations, $p_{O_2}$ the molecular oxygen pressure, and $\tilde{\mu}_{O_2}(T, p^0)$ the temperature dependent chemical potential at reference pressure $p^0$ = 1 atm. $\tilde{\mu}_{O_2}(T, p^0)$ can be determined using thermochemical tables and equals -0.54 eV at 300 K.[26,27]

**RESULTS AND DISCUSSION**

*Thin film growth* - For film growth and oxygen adsorption studies on reactive metals it is crucial to start with a thoroughly clean and ordered substrate surface, and to know how the film growth mode affects the surface morphology. We first describe our results on the growth of the Pd(111) films used in the oxygen adsorption studies. Figure 1 presents the surface morphology (a) and an AES spectrum (c) of the bulk Ru(0001) substrate surface after the cleaning procedure. The atomically flat surface morphology in STM images, the absence of impurities in the AES spectrum, and the sharp hexagonal 1×1 LEED pattern (Fig. 1(b)) without adsorbate induced reconstructions, confirm that the Ru(0001) substrate surface is clean.

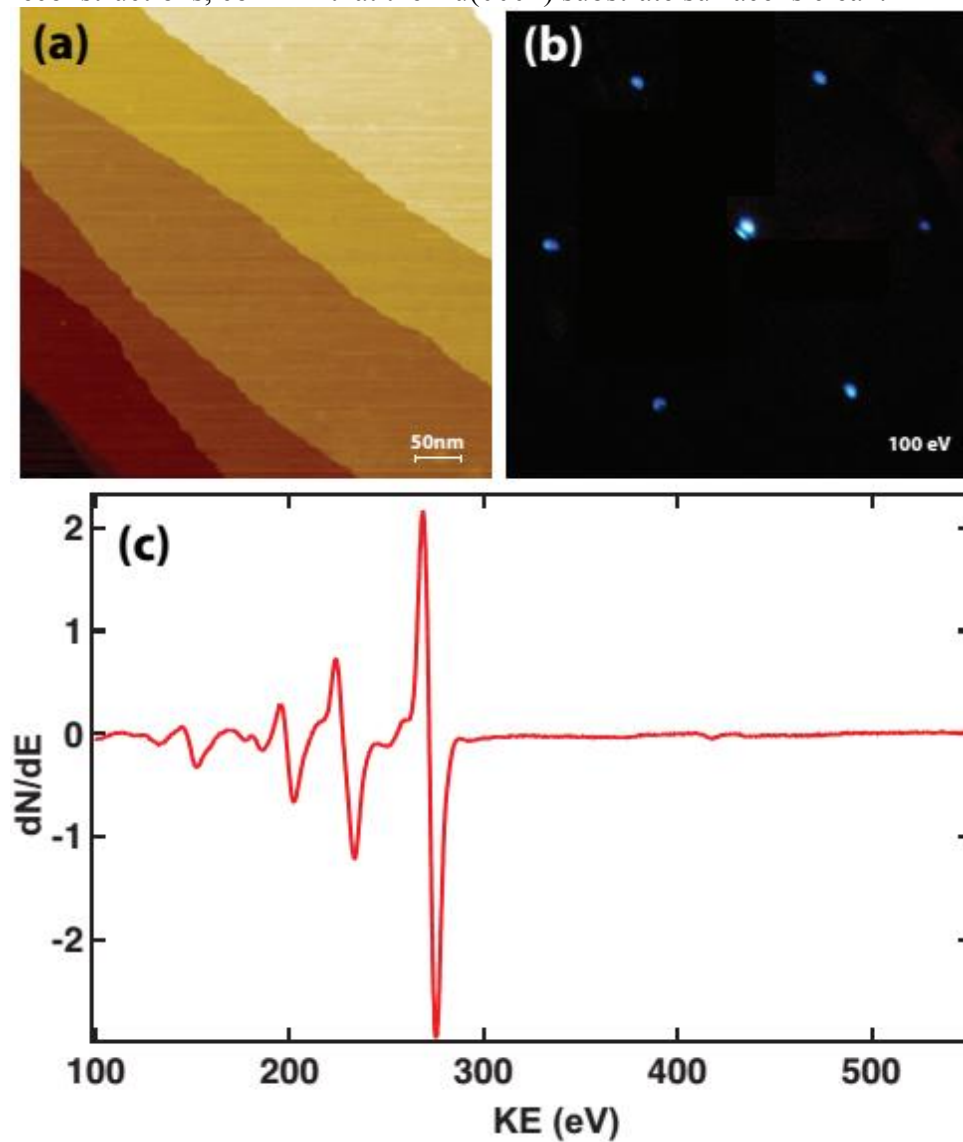

**Figure 1.** (a) 500nm×500nm STM image of the Ru(0001) surface after sputtering and annealing (see text); (b) LEED pattern of the clean Ru(0001) surface at 100 eV; (c) Ru $M_{45}VV$ Auger spectrum of the surface.

Figure 2 shows the evolution of the surface morphology with increasing Pd film thickness, grown on the Ru(0001) substrate held at room temperature. As observed in Figure 2(a), depositing 0.65 ML Pd onto a Ru surface held at room temperature results in the formation of Pd monolayer islands emanating from the Ru step edges with alternating directions due to the ABABA stacking of hcp Ru crystal, suggesting pseudomorphic film growth. As Pd atoms continue to accumulate on the surface upon further growth, the Pd islands start to expand and gradually fill the surface. These results mimic the growth of Pt on Ru(0001).[28] The Pd film does not strictly follow a layer-by-layer growth mode at room temperature; as shown in Figure 2(b) and (c) the growth of the second Pd layer commences before the first layer is completed. This trend persists up to larger film thickness (see Figure 2(d)). LEED data for different film thicknesses (not shown) maintain the 1×1 pattern already observed on the clean Ru(0001) substrate, indicating that the Pd films grow pseudomorphically and that the surface is not reconstructed.

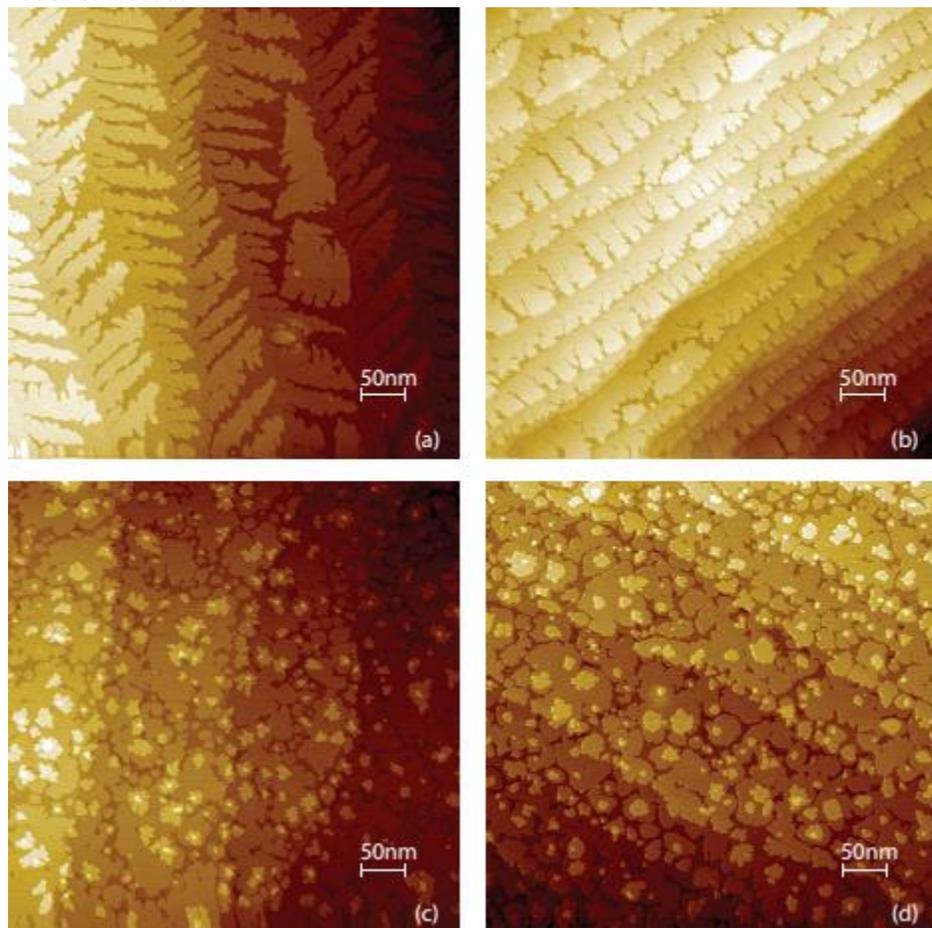

**Figure 2**. 500 nm×500 nm STM images of as-grown Pd(111) thin films with different thickness: (a) 0.65 ML; (b) 0.85 ML; (c) 1.2ML; and (d) 4.9ML.

*Oxygen adsorption* - Next we examine the chemical properties of the Pd films using oxygen adsorption experiments. We exposed the films to molecular oxygen and subsequently evaluated the quantity of adsorbed oxygen using the magnitude of the oxygen $KL_{23}L_{23}$ peak at 512 eV. To account for possible variations in beam current and sample positioning, we measured the amplitude of the oxygen $KL_{23}L_{23}$ peak relative to that of the Pd *MNN* Auger line at 330 eV. We then corrected the measured O/Pd Auger amplitude ratios for the trivial thickness dependence of the Pd Auger intensity (see Appendix B) so that the corrected variation of the O/Pd Auger amplitude ratio only reflects the thickness dependence of the surface chemical interaction with oxygen. Figure 3 presents these data, where once more we normalized the corrected amplitude ratios to that of an oxygen saturated Ru(0001) surface with an absolute coverage of 0.5 ML (see

Appendix B). The concentration of adsorbed oxygen rapidly increases and quickly saturates around 5 L for all samples, consistent with previous work.[16,18-20,29] Strikingly, the data reveal a significant non-monotonic thickness dependence of the oxygen concentration at saturation. Indeed, in Figure 4 we have plotted the oxygen coverage at 12 L as a function of the Pd film thickness, revealing a pronounced minimum in the oxygen saturation coverage, with a minimum value that is almost indistinguishable from zero.

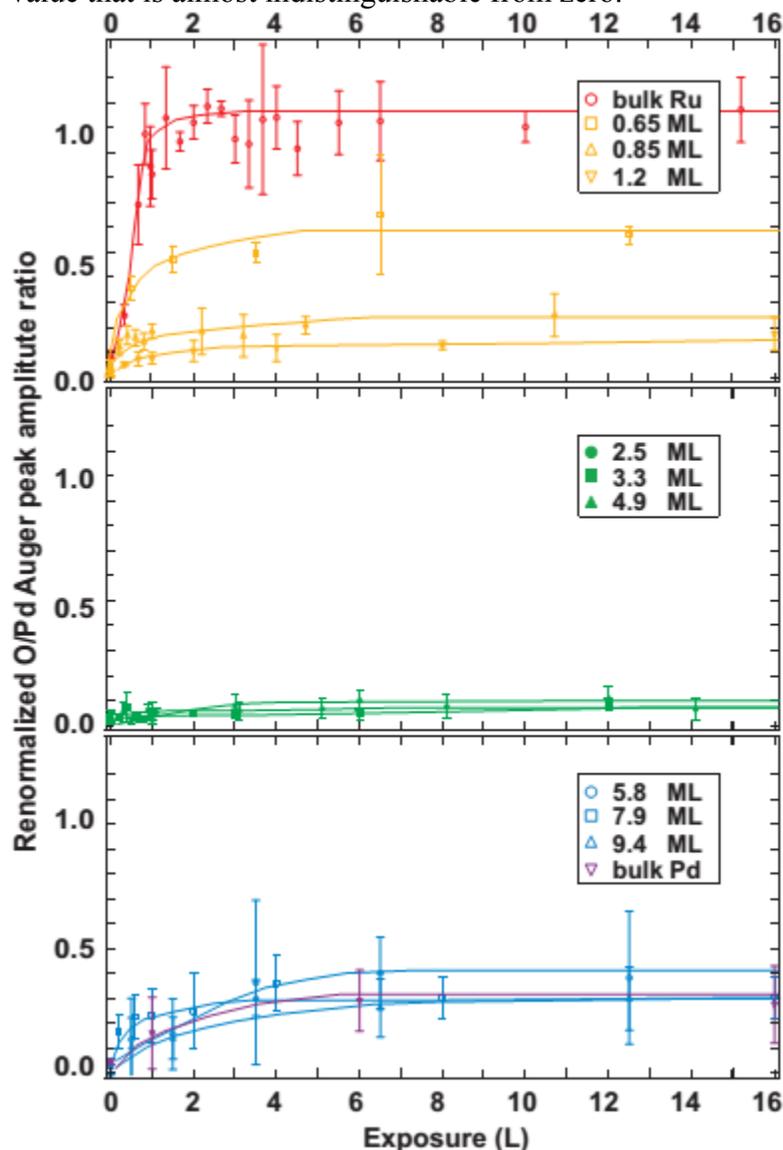

**Figure 3.** Oxygen adsorption isotherms for Pd(111) films on Ru(0001), measured with Auger electron spectroscopy at room temperature. Isotherms for bulk Ru(0001) and bulk Pd(111) are also included. Each data point in the graph represents the average intensity ratio of the oxygen $KL_{23}L_{23}$ and Pd (Ru) $M_{45}VV$ Auger signals, measured at five different locations on the sample. Error bars represent the corresponding standard deviation. The intensity ratios are corrected for the finite film thicknesses and renormalized to the calculated intensity ratio at the 0.5 ML oxygen saturation coverage on bulk Ru(0001), following the procedures in Appendix B.

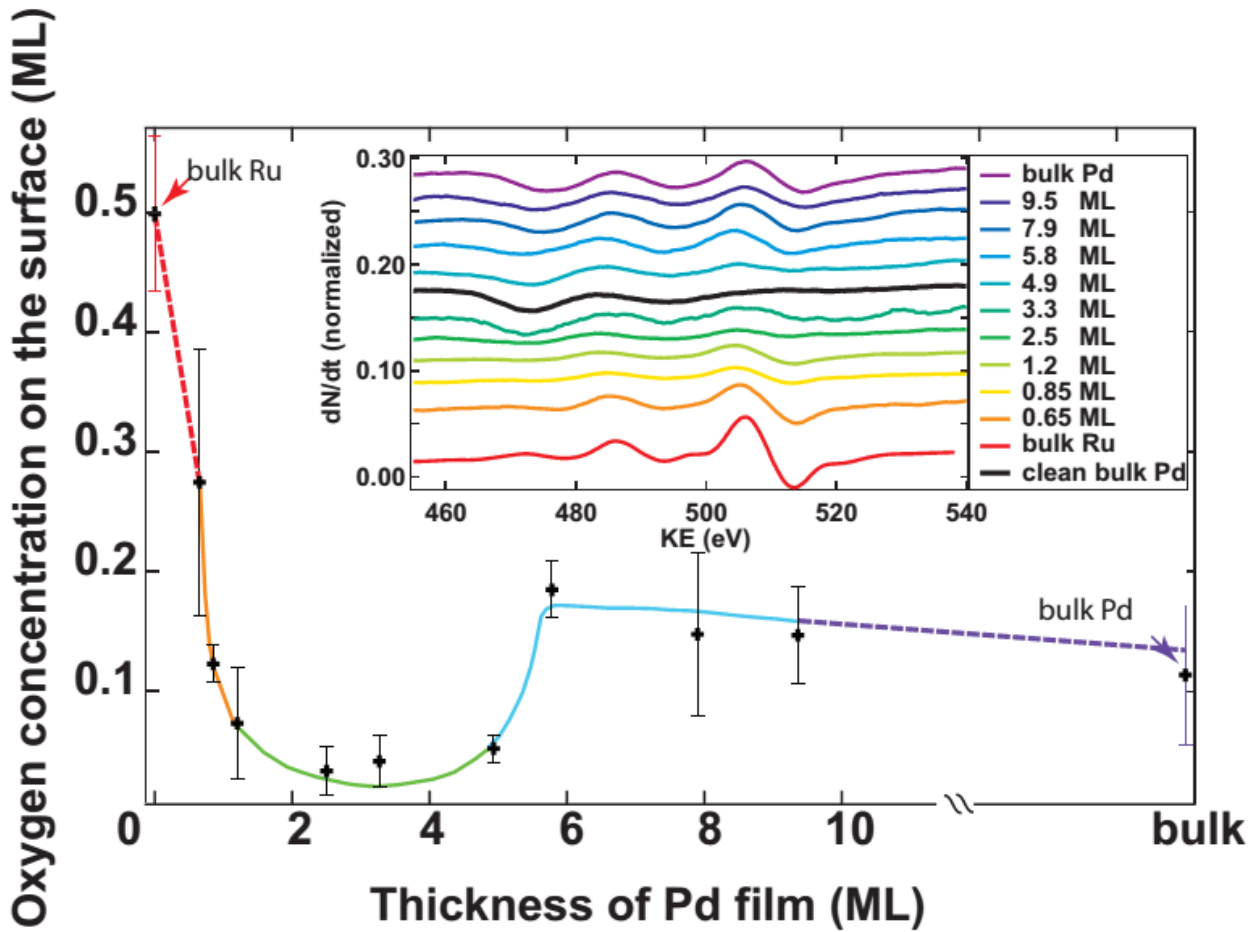

**Figure 4**. Oxygen surface concentration at the saturation coverage as a function of the film thickness. The total oxygen exposure is 12 L. Inset: Oxygen $KL_{23}L_{23}$ Auger spectra of different samples at saturation. The non-monotonic variation of the oxygen coverage is readily evident from the raw Auger spectra (inset).

This size effect can also be directly observed in the raw spectra of the oxygen $KL_{23}L_{23}$ Auger line, presented in the inset of Figure 4, showing a significant decrease in the oxygen AES amplitude for Pd film thicknesses between 2 and 5 ML. Note that the Auger spectra in this energy range also exhibit small contributions from the Pd *MNN* Auger lines. Although these Pd lines have slightly different kinetic energies, it is not possible to reliably scale and subtract them from the spectra of the oxygen exposed samples. Hence, the Pd related Auger features account for at least part of the measured oxygen intensity in the 2-5 ML Pd film thickness range, meaning that the extracted oxygen saturation coverage in Fig. 4 is an upper bound. *These ultrathin Pd films are thus surprisingly inert to oxygen exposure under the given experimental conditions.* Note also that according to the literature, the RT saturation coverage of oxygen on bulk Pd(111) is a factor two smaller than that of bulk Ru(0001).[16-18] Here, we observe that the measured AES amplitude of oxygen on bulk Pd(111) is about a factor three smaller than that of bulk Ru(0001). While this could be caused by electron-induced oxygen desorption from the Auger electron beam,[29,30] we carefully made sure that the total duration of electron beam exposure was similar for all experiments. Hence, the observation of a non-monotonic thickness dependence of the oxygen saturation coverage is not directly related to electron beam exposure. Rather, it indicates that the binding energy of oxygen on the Pd films varies non-monotonically with thickness.

*Oxygen binding energies from DFT* - In order to acquire an understanding of the surprising non-monotonic thickness dependence of the oxygen saturation coverage of these films, different mechanisms have to be considered. Apart from the presence of possible electronic size effects alluded to in the introduction, the pseudomorphic Pd films in our experiments are under compressive epitaxial strain due to the 1.5% lattice mismatch between Pd and the Ru substrate. Therefore, we first consider the influence of strain on the oxygen binding energy. Using DFT, we consider the binding energy of oxygen on a pseudomorphic (i.e., 1.5% compressed) 1 ML thick Pd(111) film on Ru(0001), and that of oxygen on a 1.5% compressively-strained 6 ML-thick freestanding Pd(111) slab. In both cases, the binding energies are lowered relative to that of oxygen on an unstrained 6 ML Pd(111) slab. In the first case, the lowering of the oxygen binding energy can be due to compressive strain, as well as other sources such as electronic size effects. In the second case, the lowering can only be due to compressive strain. Our calculations show that the binding energy lowering in the first case amounts to 630 meV (see Fig. 5) whereas it is only ~100 meV in the second case. This shows that epitaxial strain is only a minor contributor to the relative inertness of the 1ML Pd film (as well as that of the 2 and 3 ML films as the decrease in their binding energies is significantly larger than the 100 meV energy scale that is associated with the maximum compressive strain, see Fig. 5).

Next, we calculate the thickness-dependent oxygen binding energy and density of states of Pd(111) films on Ru(0001), as well as those of pure Ru(0001) and Pd(111), for the ordered 2×2 structure. These calculations provide important insights in the correlation between the density of states of the pristine film surfaces and the oxygen binding energy. Figure 5 shows the oxygen binding energies as a function of the *d*-band center of the bare surface calculated from DFT. For the Pd film samples (with the exception of the ¾ ML thick Pd film), we find that the trend follows the prediction of the Hammer-Norskov *d*-band model:[6-8,31] the binding energy increases with the decreasing *d*-band energies of the thicker Pd films and ultimately that of bulk Pd(111).[6] The difference in Ru-O binding energy is also adequately described by the Hammer-Norskov model and is mostly attributed to an enhancement in the overlap matrix. Deviations from this trend for the ¾ ML, however, are of a more subtle origin and will be discussed below.

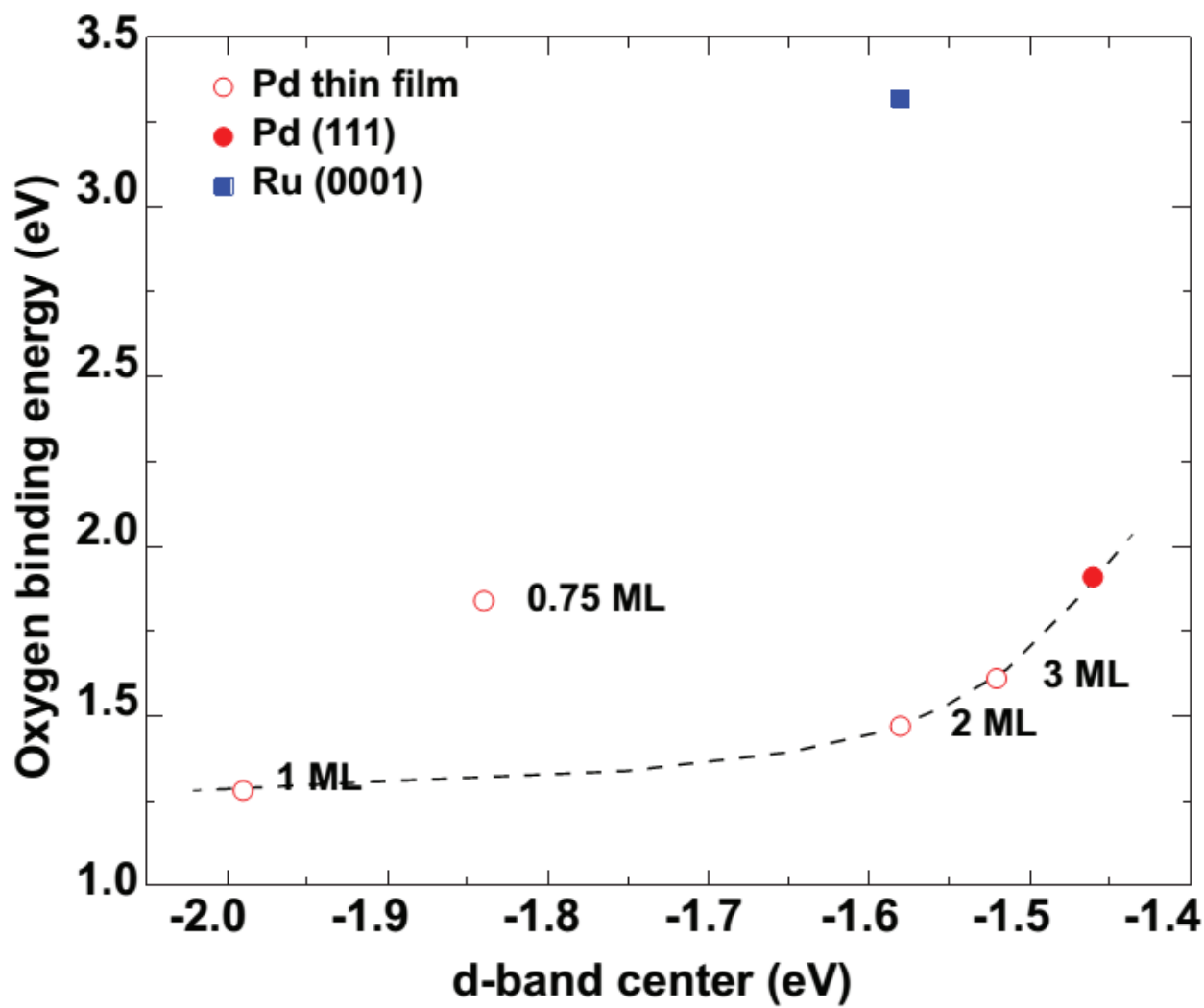

**Figure 5.** Calculated oxygen binding energies from DFT for bulk Ru(0001) and Pd(111), and for Pd(111) films on Ru(0001), plotted as function of the *d*-band center location of the pristine metal surface, relative to the Fermi energy. The dashed line serves as a guide to the eye.

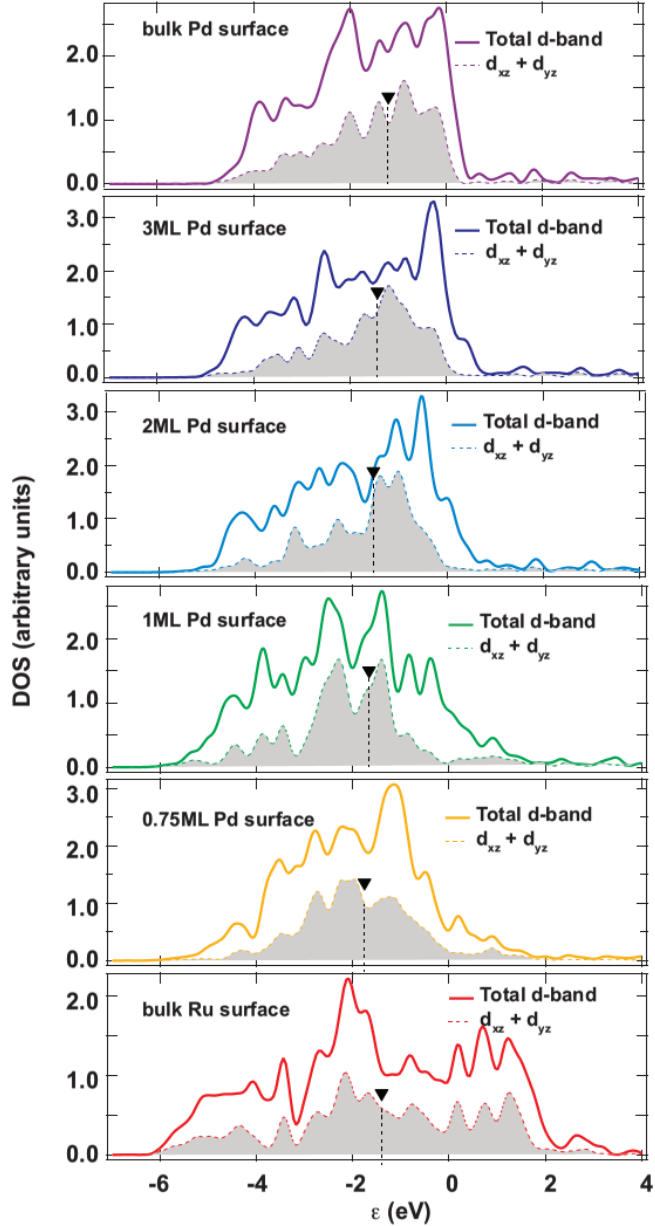

**Figure 6**. Surface projected density of states (DOS) distribution of the total *d*-band and of the $4d_{xz}$ and $4d_{yz}$ states combined (gray area) prior to oxygen adsorption. A thickness dependent shift in the centroid of total d-band is visible. The small black triangles indicate the centroids of the partial $4d_{xz}$ and $4d_{yz}$ DOS.

Charge density contours (not shown) reveal that the $4d_{xz}$ and $4d_{yz}$ orbitals are the principal participants in the oxygen-surface bonding at the fcc lattice sites, as expected. Indeed, the partial $4d_{xz}$ and $4d_{yz}$ DOS presented in Figure 6 gradually shifts towards the Fermi energy with increasing Pd film thickness. As a consequence, the antibonding O(2*p*)-Pd(4*d*) DOS feature shifts to higher energies above the Fermi level, decreasing its filling and thus increasing the oxygen binding energy; see Fig. 7.

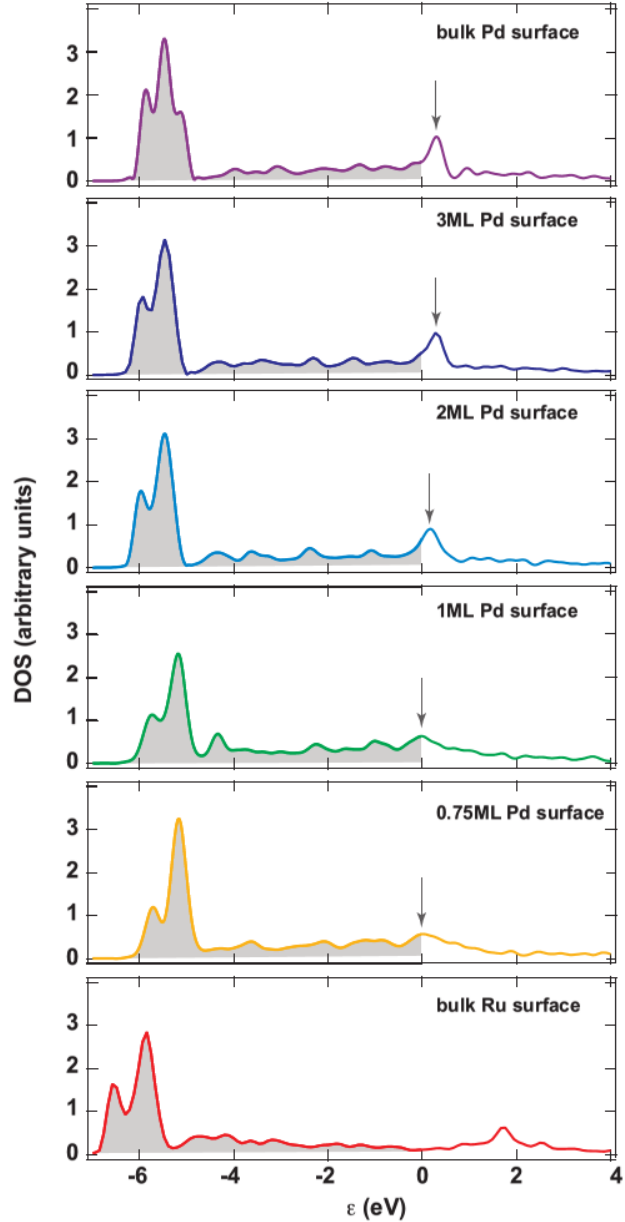

**Figure 7**. Evolution of the oxygen 2*p*-4*d* hybridized states on the surface of Pd(111)/Ru(0001) thin film samples for different film thicknesses. A shift in the 2*p*-4*d* anti-bonding states, indicated by the arrow, is clearly visible.

The shift in the 4*d*-states towards the Fermi level could in principle be the result of a charge transfer between the Pd film and the Ru substrate.[22,32] However, this effect should be insignificant for metal films with thicknesses beyond 1 ML. Indeed our DFT calculations indicate that such a charge transfer is insignificant even for the first ML of Pd, consistent with the equal electronegativity of Pd and Ru. Furthermore, we do not see evidence of a clear quantum size effect in the calculated electronic structure (such as the existence of well-separated 2D subbands). Hence, we conclude that the shift of the $4d_{xz}$ and $4d_{yz}$ states towards the Fermi energy with increased film thickness originates from finite size effects on the out-of-plane metal bonding.

As observed above in Fig. 5, the ¾ ML Pd film surface defies this trend. In particular, the oxygen binding energy is significantly higher than would be expected based on the location of the *d*-band center, orbital filling or overlap matrix. Here, as a consequence of the lower Pd density, the oxygen atoms relax deeper into the surface layer, and the *d*-orbitals involved in the metal-oxygen bonds in the hollow site are now dominated by the narrower in-plane $d_{x2-y2}$ orbitals. Fitting our results to the *d*-band model,[33] we find that the *s-d* coupling matrix elements, $V_{sd}$, should be significantly larger for both the ¾ ML thick Pd film and the Ru(0001) surface, as compared to those of the Pd(111) films and bulk Pd(111) surface, thus leading to an enhancement of the respective oxygen binding energies. These results indicate that the general *d*-band center argument is valid only when the adsorption site and orbital symmetries remain the same. Our results emphasize the importance of understanding the nature of the orbitals involved in adsorption and catalysis.[32,34]

*Adsorption energy* - Finally, we calculate the Gibbs free energy of adsorption per 2×2 unit cell for different film thicknesses as a function of the oxygen chemical potential at 300 K according to Eq. (1),[26] see Figure 8. Here, we included the results for both ¼ ML and ½ ML oxygen coverages. The lines with a steep slope correspond to an oxygen coverage of ½ ML while those with the gentle slope correspond to ¼ ML coverage. In this plot, the zero of the chemical potential corresponds to half the total energy of a free oxygen molecule (4.88 eV) calculated with spin polarized DFT. In other words, the variation in chemical potential along the horizontal axis only includes the pressure and temperature dependent terms in Eq. 2, which are determined by the experimental conditions. The corresponding pressure scale is indicated at the top. The experimentally relevant pressures are the pressures during oxygen exposure ($p≈10^{-8}$ mbar; $p_{O_2}≈10^{-11}$ atm) and Auger measurement ($p≈10^{-10}$ mbar; $p_{O_2}≈10^{-15}$ atm), where the last estimate is of course very crude. Accordingly, the experimental chemical potential is estimated to be in the range of -0.6 to -0.7 eV.

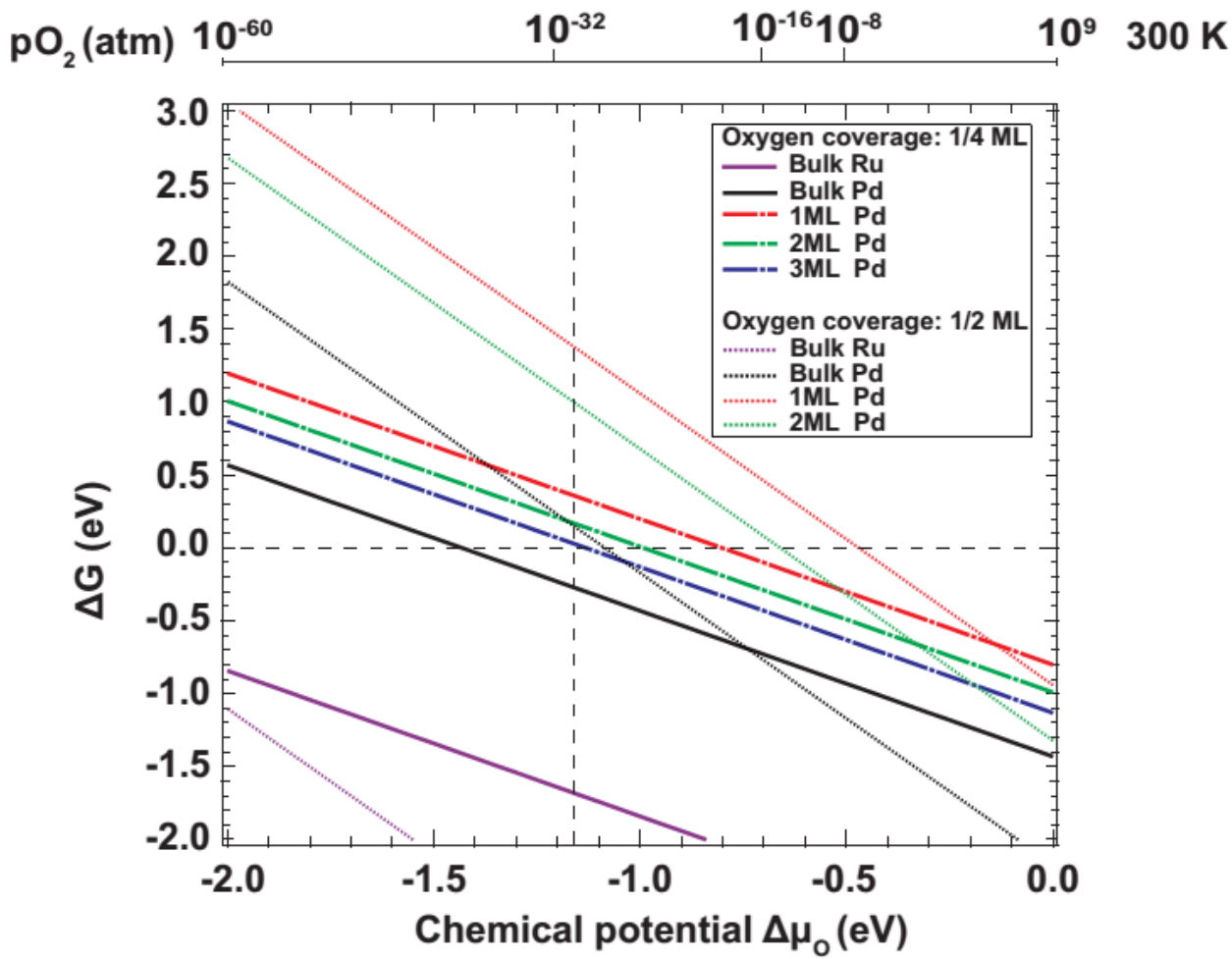

**Figure 8**. Change in the Gibbs free energy $\Delta G$ per 2×2 unit cell upon oxygen adsorption on bulk Pd(111), Ru(0001), and on thin Pd(111) films as a function of the oxygen chemical potential $\Delta\mu_O$ at room temperature relative to half the total energy of a free oxygen molecule. Negative values of $\Delta G$ indicate that adsorption is thermodynamically favorable.

Fig. 8 reproduces the experimental trends quite well. Clearly, oxygen binds very strongly with the Ru(0001) surface and the ½ ML phase is strongly favored over the ¼ ML phase for the entire range of chemical potentials on Ru(0001). For bulk Pd(111), these phases are very close in energy for -0.7 eV < $\Delta\mu$ < -0.6 eV but the adsorption energy is greatly reduced. The ¼ ML phase clearly prevails for the Pd films but the adsorption energies have become very small under the experimental conditions. The latter is qualitatively consistent with the lack of adsorbed oxygen in the Auger measurement. Full quantitative agreement should not be expected in light of the DFT errors,[35] crude estimates of the chemical potentials, and the exclusion of kinetic considerations. The latter are beyond the scope of the investigations. Nonetheless, we believe that the present comparison of theory and experiment provides a compelling validation of the *d*-band model in a clean approach where only the energy of the *d*-band center is varied.

Lastly, we would like to note that the reported presence of subsurface oxygen[19,20,29,36,37] on bulk Pd(111) does not affect the general validity of our conclusions. While subsurface oxygen would affect our estimate of the surface oxygen concentration from AES, the experimental results in Refs. [19,20,29,36,37] consistently reveal that subsurface oxygen always requires significant amounts of atop oxygen, which would easily be detected by AES. Therefore, the remarkable inertness of the ultrathin Pd films toward oxidation remains firmly established.

**SUMMARY AND CONCLUSIONS**

In conclusion, we have studied oxygen adsorption on ultrathin pseudomorphic Pd(111) films on Ru(0001). The oxygen saturation coverage at room temperature evolves non-monotonically as a function of the film thickness, exhibiting a pronounced minimum between 2 and 5 ML of Pd where the film surface is remarkably inert toward oxygen adsorption. Aided by DFT calculations, our analysis suggests that a finite size effect is the cause of these surprising observations: the $4d_{xz}$ and $4d_{yz}$ orbitals involved in the metal-oxygen bonds become increasingly delocalized with increasing film thickness, while their centers of gravity shifts toward the Fermi energy. This, in turn, increases the oxygen binding energy on the surface conform the $d$-band model. The ¾ ML Pd film deviates from this trend because the symmetry of the orbitals involved in the metal-oxygen bonding is different.

Our focus on pseudomorphic metal films allowed us to specifically test the validity of the $d$-band model without additional complications arising from the presence of catalytically active sites associated with the presence of e.g., foreign species, reduced coordination, or atomic steps on the surface. Specifically, we were able to establish a clear correlation between oxygen adsorption and surface electronic structure. The $d$-band center location seems a powerful predictor for establishing chemical trends in heterogeneous catalysis, provided that the adsorption sites and bonding orbital symmetries remain unchanged.

**APPENDIX A**

Under the chosen growth conditions, Pd films grow approximately layer-by-layer on Ru(0001).[38] The growth rate can be calibrated by measuring the intensity of a Pd Auger line as a function of deposition time.[39] However, as the Pd flux changes over time, it is essential to have a reliable verification of the Pd film thickness following sample growth. This was done as follows.

Let the Pd $M_{45}VV$ Auger intensity (at 330 eV) of a monatomic Pd layer be denoted by $I_1$. Likewise, let the Ru $M_{45}VV$ Auger intensity (230 eV) of a monatomic Ru layer be denoted $I_2$. Then, the ratio of the Pd and Ru Auger signals for a thin film system consisting of $n$ complete monolayers of Pd on bulk Ru is given by

$$\frac{I_{Pd}^n}{I_{Ru}^\infty} = \frac{I_1}{I_2} \times \frac{\frac{1-\exp[-na_1/\lambda_{11}]}{1-\exp[-a_1/\lambda_{11}]}}{\frac{\exp[-na_1/\lambda_{21}]}{1-\exp[-a_2/\lambda_{22}]}} \equiv \frac{I_1}{I_2} \times \frac{I_1^n}{I_2^n} \qquad \text{A.1}$$

where it is assumed that the Auger intensity of a monatomic layer is attenuated exponentially by the layers above due to inelastic scattering of the Auger electrons.[39] Here, $a_1$ and $a_2$ are the lattice constants of Pd and Ru along the growth direction, respectively; $\lambda_{11}$ is the inelastic mean free path of the Pd Auger electrons in the Pd film, and $\lambda_{22}$ and $\lambda_{21}$ are the inelastic mean free paths of the Ru Auger electrons in bulk Ru and in the Pd overlayer, respectively. $I_1^n$ and $I_2^n$ are defined by the corresponding numerator and denominator in the middle expression of Eq. (1).

For non-integer layer thickness ($n+f$) with $0 < f < 1$, the corresponding intensity ratio can be written as

$$\frac{I_{Pd}^{n+f}}{I_{Ru}^{\infty}} = \frac{I_1}{I_2} \times \frac{fI_1^{n+1} + (1-f)I_1^n}{fI_2^{n+1} + (1-f)I_2^n} \quad \text{A.2}$$

The only unknown parameter in above expressions is $I_1/I_2$. This parameter is determined by growing a sub-monolayer Pd film with $n = 0$ and f representing the fractional film coverage. The latter can be determined by STM so that the ratio $I_1/I_2$ can be calculated from the corresponding Auger intensity ratio. The right hand side of equation A2 is then calculated and tabulated for a range of values of ($n, f$) and the film thickness can be read off from the experimental intensity ratio. Note that the use of intensity ratios as opposed to absolute Auger intensities also alleviates uncertainties due to, *e.g.*, fluctuations in Auger beam current and sample positioning between successive experiments.

**APPENDIX B**

The oxygen coverage on the Pd films is represented by the ratio of the oxygen $KL_{23}L_{23}$ and Pd $M_{45}VV$ Auger intensities. Because the Pd Auger intensity scales with the number of atom layers in the Pd film, we renormalized the Pd Auger intensity to that of bulk Pd(111) so as to obtain a O($KL_{23}L_{23}$)/Pd($M_{45}VV$) intensity ratio that is only determined by the surface chemistry. The renormalization is based on the assumption that

$$I_{Pd}^n = I_1 \times \frac{1 - \exp[-na_1/\lambda_{11}]}{1 - \exp[-a_1/\lambda_{11}]} \quad \text{B.1}$$

and

$$I_{Pd}^{\infty} = I_1 \times \frac{1}{1 - \exp[-a_1/\lambda_{11}]} \quad \text{B.2}$$

meaning that the experimental O($KL_{23}L_{23}$)/Pd($M_{45}VV$) intensity ratio needs to be renormalized by a factor $1 - \exp[-na_1/\lambda_{11}]$.

In order to relate these renormalized intensity ratios to an absolute coverage scale, we measured the O($KL_{23}L_{23}$)/Ru($M_{45}VV$) ratio on an oxygen saturated Ru(0001) surface with precisely known coverage (0.5 ML at room temperature). To have a meaningful comparison between the renormalized O($KL_{23}L_{23}$)/Pd($M_{45}VV$) and O($KL_{23}L_{23}$)/Ru($M_{45}VV$) Auger intensity ratios, we rescaled the measured Ru Auger intensity using

$$\frac{I_{Pd}^{\infty}}{I_{Ru}^{\infty}} = \frac{I_1}{I_2} \times \frac{1 - \exp[-a_2/\lambda_{22}]}{1 - \exp[-a_1/\lambda_{11}]} \quad \text{B.3}$$

**Author Contributions**
The manuscript was written through contributions of all authors. All authors have given approval to the final version of the manuscript.

## ACKNOWLEDGMENT

This research was supported by the U.S. Department of Energy, Office of Science, Basic Energy Sciences, Materials Sciences and Engineering Division. This research used resources of the National Energy Research Scientific Computing Center, which is supported by the DOE Office of Science under Contract No. DE-AC02-05CH11231.